\documentclass[prl, twocolumn, showpacs, floatfix]{revtex4}  
\pdfoutput=1
\usepackage[hang]{footmisc}
\usepackage{hyperref}
\usepackage{amssymb}
\usepackage{amsmath}
\usepackage{float}
\usepackage{bbm}
\usepackage{graphicx}
\usepackage{epsfig}
\usepackage{epstopdf}
\usepackage{verbatim}
\usepackage[usenames]{color}

\begin{document}
\title{Enhancement of condensate depletion due to spin-orbit coupling}
\author{Xiaoling Cui$^{1,2}$ and Qi Zhou$^{3}$}
\affiliation{$^{1}$ Institute for Advanced Study, Tsinghua University, Beijing 100084, China\\
$^{2}$ Department of Physics, The Ohio State University, Columbus, OH 43210, USA\\
$^{3}$ Department of Physics, The Chinese University of Hong Kong,
Shatin, New Territories, Hong Kong}
\date{{\small \today}}

\begin{abstract}
We show that spin-orbit coupling(SOC) significantly enhances the depletion of a homogeneous Bose-Einstein condensate in three dimension. With decreasing anisotropy of SOC, both quantum and thermal depletion increase. Particularly, different types of SOC give rise to qualitatively different dependences of condensate depletion on microscopic variables including  scattering length, strength of SOC and temperature, a novel feature that can be directly observed once these types of SOC are realized in experiments. Moreover, we point out that  thermal depletion in three dimension becomes logarithmically divergent at any given finite temperature when both SOC and interaction approach the isotropic limit.  
\end{abstract}
\maketitle

The recent realization of synthetic spin-orbit coupling(SOC) for ultra cold atoms 
is an exciting development in the field of quantum gases\cite{NIST, Chen, Zhang, Martin}.  As many experimental parameters, including density, interaction and the configuration of SOC itself\cite{proposal},  are highly controllable, quantum gases provides physicists an ideal platform to
investigate the interplay between SOC and interaction in many-body systems, a
challenging problem that remains unsolved in condensate matter
physics so far \cite{TI}.

In the literature, extensive studies have predicted a number of novel properties of Bose-Einstein condensate in the presence of SOC\cite{Zhai, Galitski,NIST_type,Wu,Rashba_trap,Goldbart, Ozawa1,Barnett, Ozawa, Hu}. However, condensate depletion, an intrinsic property of bosons, has not been systematically discussed except for a few numerical studies for Rashba SOC\cite{Ozawa, Barnett}. In the absence of SOC, one important effect of condensate depletion is that it destroys a condensate at low dimensions. In three dimension, the depletion for weakly interacting bosons is small and a condensate naturally exists at low temperatures\cite{Pethick}.  However, in the presence of SOC, due to the lack of a systematical study of analytic expressions for the condensate depletion, it is unclear how the depletion depends on microscopic parameters, and in particular, on the configuration of SOC.
In this Rapid Communication, we point out that condensate depletion becomes significant even in three dimension in the presence of SOC and may completely destroy the condensate at low temperatures. We also obtain  the explicit forms for the dependence of condensate depletion on the anisotropy and strength of SOC, the scattering length, and the temperature.


The single-particle Hamiltonian with a general SOC between two spin states $|\uparrow\rangle$ and $|\downarrow\rangle$ can be written as
\begin{equation}
\mathcal{H}_0=\frac{1}{2m}\Big\{ \sum_{\alpha}(-\partial_{\alpha}^2-2i\kappa_{\alpha}\sigma_{\alpha}\partial_{\alpha}) + \kappa_x^2 \Big\}.
\end{equation}
where $\kappa_x\ge \kappa_y,\kappa_z$ has been assumed without loss of generality. Throughout the paper we set $\hbar=1$. $\kappa_\alpha$ is the strength of SOC, and $\sigma_{\alpha}(\alpha=x,y,z)$ are Pauli matrices. The eigen-state has
two branches $|{\bf k}^{+}\rangle=u_{\bf k}|{\bf
k}_{\uparrow}\rangle +v_{\bf k} e^{i\phi_{\mathbf{k}}}|{\bf
k}_{\downarrow}\rangle$, $|{\bf k}^{-}\rangle=-v_{\bf
k}e^{-i\phi_{\mathbf{k}}}|{\bf k}_{\uparrow}\rangle +u_{\bf k} |{\bf
k}_{\downarrow}\rangle$, 
corresponding to eigen-energy
\begin{equation}
\epsilon_{{\bf k}}^{\pm}=\frac{1}{2m}\left(k_x^2+k_y^2+k_z^2\pm
2 A_{\boldsymbol \kappa, {\bf k}}+\kappa_x^2 \right)\label{ek},
\end{equation}
with $A_{\boldsymbol \kappa, {\bf k}}=\sqrt{\kappa_x^2k_x^2+\kappa_y^2k_y^2+\kappa_z^2k_z^2}$, $u_{\bf k}=\sqrt{\frac{1}{2}(1+\frac{\kappa_z k_z}{A_{\boldsymbol \kappa, {\bf k}}})}$,
$v_{\bf k}=\sqrt{\frac{1}{2}(1-\frac{\kappa_z k_z}{A_{\boldsymbol \kappa, {\bf k}}})}$, and 
$\phi_{\mathbf{k}}=arg(\kappa_x k_x+i\kappa_y k_y)$.

For isotropic SOC where $\kappa_x=\kappa_y=\kappa_z=\kappa$, Eq.(\ref{ek}) becomes $\epsilon_{{\bf k}}^{\pm}=\frac{1}{2m}(|{\bf k}|\pm \kappa)^2$. 
 The lower branch ($\epsilon_{{\bf k}}^{-}$) gives rise to the single particle density of states (DOS) $\rho_0(\epsilon)\sim1/\sqrt{\epsilon}$ for $\epsilon\rightarrow 0$\cite{Brazovskii, Vijay1}, the same as that for one dimensional systems in the absence of SOC. As condensate depletion of non-interacting bosons solely relies on single-particle DOS,  we conclude {\it a non-interacting condensate in three dimension is completely destroyed by the isotropic SOC even at zero temperature}. 
The same conclusion applies to Rashba coupling in two dimension, where $\kappa_x=\kappa_y$. 

\begin{figure*}
\includegraphics[width=5.6in]{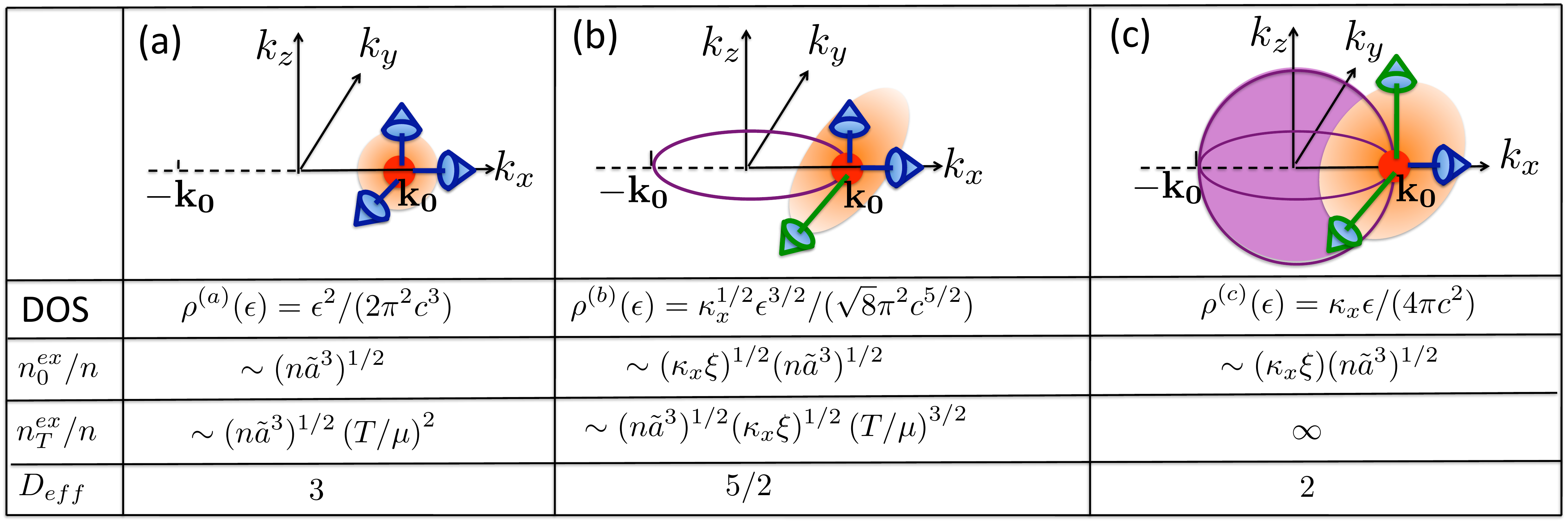}
\caption{(Color Online) DOS and condensate depletion for three particular configurations of SOC. Red spheres represent the condensates, and orange clouds represent the depletion, which is enhanced along the $y$ direction in (b) and the $y$ and $z$ directions in (c). Purple circle in (b) and sphere in (c) represent the kinetic energy minimums. Short blue arrows represent the sound modes and long green arrows represent the soft modes.  
The effective dimension ($D_{eff}$) is obtained by comparing DOS with that for a $D_{eff}$ dimensional system without SOC. 
}
\end{figure*}

While the above conclusion on condensate depletion of non-interacting bosons is unambiguous, it is essential to ask what are the condensate depletion in the presence of both SOC and interaction, and how the depletion depends on the anisotropy of SOC, which always exists in practice.  We consider the interaction between spin-orbit coupled bosons in the mean-field region\cite{RG}, $H_{int}=\int d^3r(2U_{\uparrow\downarrow}\hat{n}_{\uparrow}({\bf r})\hat{n}_{\downarrow}({\bf r})+U\sum_{\sigma}\hat{n}^2_\sigma(\bf r))$, where $\hat{n}_\sigma({\bf r})$ is the density operator for spin $\sigma(=\uparrow,\downarrow)$ and  $U_{\sigma\sigma'}$ is the interaction strength. For weak interactions, chemical potential is much smaller than the energy difference
between the bottoms of two branches, i.e., $\mu\ll
E_{\kappa}\equiv\kappa_x^2/2m$,  the negligible occupation in the higher branch justifies that $H_{int}$ can be expanded in the lower branch $(a^{\dag}_{{\bf k}}, a_{{\bf k}})$, as 
\begin{equation}
H_{int}=\frac{1}{2\Omega}\sum_{{\bf k_1,k_2,k_3,k_4}}f_{{\bf k_1,k_2}}^{\bf k_3,k_4} a^{\dag}_{\bf
k_1}a^{\dag}_{{\bf k_2}}a_{{\bf
k_3}}a_{{\bf k_4}}\delta_{\bf k_1+k_2-k_3-k_4}, 
\end{equation}
where
\begin{equation}
\begin{split}
f_{\bf k_1,k_2}^{\bf k_3,k_4} &={U} v_{{\bf k_1}}v_{{\bf k_2}}v_{{\bf k_3}}v_{{\bf
k_4}}e^{i(\phi_{{\bf k_1}}+\phi_{{\bf k_2}}-\phi_{\bf k_3}-\phi_{\bf k_4})}\\
&+{U}u_{{\bf
k_1}}u_{{\bf k_2}}u_{{\bf
k_3}}u_{{\bf
k_4}}\\
&+2{U_{\uparrow\downarrow}}v_{\bf
k_1}u_{\bf k_2}u_{\bf
k_3}v_{\bf k_4}e^{i( \phi_{{\bf
k_1}}-\phi_{{\bf k_4}})},
\end{split}
\end{equation}
and $\Omega$ is the volume. $U_{\uparrow\uparrow}=U_{\downarrow\downarrow}\equiv U$ has been assumed to simplify notations.

{\bf Bogoliubov spectrum} It has been shown that the mean field ground state in two dimension can be a plane-wave condensate or a stripe phase\cite{Zhai}. In three dimension, the situation is similar but more sophisticated \cite{supple}.  In this Rapid Communication,  we concrete on the plane-wave condensate 
 at $\mathbf{k}_0=\kappa_x\mathbf{e}_x$, the mean-field ground state when $U\ge U_{\uparrow\downarrow}$ and $\kappa_x\geq \kappa_y,\ \kappa_z$. 

We perform Bogoliubov theory to study the quasi-particle spectrum, which can be written as
$E_{\pm, {\bf q}}=E_{\bf q}\pm ({\epsilon_{+,{\bf
q}}-\epsilon_{-,{\bf q}}})/{2}$, where 
\begin{equation}
E_{\bf
q}=\sqrt{({\epsilon_{+,{\bf q}}+\epsilon_{-,{\bf
q}}})^2/4-V_{{\bf q}}^2}\label{ES},
\end{equation} 
and  $\epsilon_{\pm,{\bf q}}=\epsilon_{\bf k_0\pm q}+\frac{n}{2}(U+2U_{\uparrow\downarrow}u_{\bf k_0\pm q}v_{\bf k_0\pm q}\cos\phi_{\bf k_0\pm q})$, $V_{{\bf q}}=\frac{n}{2}\big[Uv_{\bf k_0+q}v_{\bf k_0-q} e^{i(\phi_{\bf k_0+q}+\phi_{\bf k_0-q})}+Uu_{\bf k_0+q}u_{\bf k_0-q}+U_{\uparrow\downarrow}(v_{\bf k_0+q}u_{\bf k_0-q}e^{i\phi_{\bf k_0+q}}+u_{\bf k_0+q}v_{\bf k_0-q}e^{i\phi_{\bf k_0-q}})\big]$. For our discussions, it is useful to enumerate $E_{\bf q}$ for three particular configurations of SOC, (a) $\kappa_x\neq 0$,  $\kappa_y=\kappa_z=0$,
(b) $\kappa_x=\kappa_y, \kappa_z=0$,
(c)  $\kappa_x=\kappa_y=\kappa_z, U=U_{\uparrow\downarrow}$. (a) is the case when SOC  exists along only one direction, (b) is the Rashba coupling, and (c) is referred to as isotropic SOC and interaction.  For these three cases, $E_{\bf q}$ in low momentum limit have distinct expressions, 
\begin{equation}
\begin{tabular}{ll}
(a): \,\,\, \,\,&$E_{\bf q}=\sqrt{\frac{\mu}{m}(q_x^2+q_y^2+q_z^2)}$\\
(b):\,\,\, \,\, &$E_{\bf q}=\sqrt{\frac{\mu}{m}(q_x^2+\frac{q_y^4}{4\kappa_x^2}+q_z^2)}$ \\
(c): \,\,\, \,\,&$E_{\bf q}=\sqrt{\frac{\mu}{m}\left(q_x^2+\frac{(q_y^2+q_z^2)^2}{4\kappa_x^2}\right)}$,\label{qen}
\end{tabular}
\end{equation}
where {\bf $q=|{\bf q}|$}, the chemical potential $\mu=n\tilde{U}$, and $\tilde{U}=(U+U_{\uparrow\downarrow})/2$. Since $\epsilon_{+,{\bf
q}}-\epsilon_{-,{\bf q}}\sim o(q^3)$, $E_{\bf q}$ represents the quasi-particle spectra in the long wavelength limit. For case (a),  {\bf $E_{\bf q}$} is linear along all three directions and characterized by an isotropic sound velocity, i.e., $E_{\bf q}=cq$, where $c=\sqrt{\mu/m}$ the same as the sound velocity for an ordinary condensate without SOC. For Rashba coupling in case (b), the dispersion becomes quadratic along the $y$ direction, i.e., a soft mode emerges(see also \cite{Wu,Barnett,Ozawa}). In case (c), the second soft mode emerges along the $z$ direction,  and this requires that both SOC and interaction are isotropic\cite{supple}. These soft modes have profound effects on condensate depletion.  

{\bf Depletion and DOS of Quasi-particles} In the framework of Bogoliubov theory, condensate depletion is given by
$n^{ex}=n^{ex}_0+n^{ex}_T$, where
\begin{eqnarray}
&n^{ex}_0&=\frac{1}{2\Omega} \sum_{\bf q}  (\frac{\epsilon_{+,{\bf q}}+\epsilon_{-,{\bf q}}}{2 E_{\bf q}}-1), \label{nex_T0} \\
&n^{ex}_{T}&=\frac{1}{2\Omega} \sum_{\bf q} \frac{\epsilon_{+,{\bf
q}}+\epsilon_{-,{\bf q}}}{2 E_{\bf q}}\left(\frac{1}{e^{\frac{E_{+,{\bf
q}}}{T}}-1}+\frac{1}{e^{\frac{E_{-,{\bf q}}}{T}}-1}\right). \label{nex_T}
\end{eqnarray}
$n^{ex}_0$ and  $n^{ex}_T$ are quantum and thermal depletion respectively. It is interesting to note that for case (a), where $u_{\bf k}=v_{\bf k}= 1/\sqrt{2}$ and $\phi_{\bf k}=0$,  Eqs.(\ref{nex_T0},\ref{nex_T}) reduce to the ones for an ordinary condensate without SOC, where  $n_0^{ex}/n\sim (n\tilde{a}^3)^{1/2}$ and $n_T^{ex}/n\sim (n\tilde{a}^3)^{1/2}(T/\mu)^2$ for $T<\mu$\cite{Pethick}. This fact allows a direct comparison among cases with or without SOC, as discussed later.

Both the quantum and thermal depletion can be written as functions of a few dimensionless quantities, 
\begin{eqnarray}
{n^{ex}_0}/{n}&=&F_0(n\tilde{a}^3, \kappa_{\alpha}\xi,\,\, ({U-U_{\uparrow\downarrow}})/{\tilde{U}}),\\
{n^{ex}_T}/{n}&=& F_T(n\tilde{a}^3, \kappa_{\alpha}\xi, ({U-U_{\uparrow\downarrow}})/{\tilde{U}}, {T}/{\mu}) \label{sca}, 
\end{eqnarray}
where $\tilde{a}=m\tilde{U}/(4\pi)$ is the averaged scattering length, and  $\xi=(2mn\tilde{U})^{-1/2}$ is the healing length.  Whereas the values of $F_0$ and $F_T$ can be calculated numerically, it is useful to obtain an analytical expression of them using a qualitative analysis as follows.  For quantum depletion in Eq.(\ref{nex_T0}), the contribution from $E_{\bf q}>\mu$ to the integral is found to be rather small, similar to the case without SOC\cite{Pethick}.  In the low energy regime, we further approximate $(\epsilon_{+,{\bf q}}+\epsilon_{-,{\bf q}})/2$ by $\mu$, its asymptotic value when $q\rightarrow 0$. Then Eq.(\ref{nex_T0}) is reduced to 
\begin{equation}
n^{ex}_0\approx \mu \int_0^\mu d\epsilon \rho(\epsilon)/(2\epsilon),\label{an0}
\end{equation}
where $\rho(\epsilon)$ is the low energy DOS of quasi-particles determined by Eq.(\ref{ES}). 
Similarly, for thermal depletion in Eq.(\ref{nex_T}), the dominated contribution to the integral comes from $E_{\bf q}<T$, and Eq.(\ref{nex_T}) can be approximated by
\begin{equation}
n^{ex}_T\approx \mu T\int_0^T d\epsilon \rho(\epsilon)/\epsilon^2. \label{anT}
\end{equation}
for low temperatures $T<\mu$\cite{footnote}. Eqs.(\ref{an0}, \ref{anT}) clearly demonstrate the important effect of $\rho(\epsilon)$ on the condensate depletion.

Using Eq.(\ref{qen}), DOS of quasi-particles at low $\epsilon$ for cases (a-c), which are represented by $\rho^{(a)}(\epsilon)$, $\rho^{(b)}(\epsilon)$ and $\rho^{(c)}(\epsilon)$ respectively,  can be obtained as shown in Table I. It is also useful to define an effective dimension according to the power-law behavior of $\rho(\epsilon)$ . Considering a $D$-dimension system without SOC, the phonon-like quasi-particle ($E_{\bf q}=cq$) gives rise to DOS as $\rho(\epsilon)\sim \epsilon^{D-1}$.  As a result,  $\rho^{(a)}(\epsilon)$, $\rho^{(b)}(\epsilon)$ and $\rho^{(c)}(\epsilon)$ correspond to effective dimensions $D_{eff}=3, 5/2$ and $2$ respectively. The reduced effective dimensions in cases (b) and (c) strongly indicates that the  condensate depletion is significantly enhanced.

 Substituting the corresponding $\rho(\epsilon)$ to Eqs.(\ref{an0}, \ref{anT}), we obtain the analytical expressions for condensate depletion in cases (a-c). As shown in the Tabel in Fig.1, both $F_0$ and $F_T$ take the form of scaling functions of a few microscopic variables.  Two important results can be revealed from these forms.

(i) Condensate depletion is largely enhanced in (b) and (c), as expected from the observation that $D_{eff}$ is reduced in these two cases.  Compared with case (a),  $n^{ex}_0/n$ is enhanced by a factor of $(\kappa_x{\xi})^{1/2}$ for (b) and a factor of $\kappa_x{\xi}$ for (c). 
Note that here $\kappa_x{\xi}=\sqrt{E_{\kappa}/\mu}\gg 1$.
If $\kappa_x$ continuously increases, there is essentially no upper limit for the enhancement. In current experiments, $\kappa_x\sim 1/\lambda_L$\cite{NIST,Chen, Zhang, Martin}, where $\lambda_L$ is the wavelength of the Raman field, and $\kappa_x\xi\sim \xi/\lambda_L\approx 5-10$. Therefore, $n^{ex}_0/n$ increases severalfold in case (b) and (c). For thermal depletion, the dependence of $n_T^{ex}/n$ on temperature becomes  $\sim(T/\mu)^{3/2}$ in case (b), in addition to the enhancement factor $(\kappa_x{\xi})^{1/2}$. Compared with the result $\sim (T/\mu)^{2} $ in case (a), this means an even larger thermal depletion in case (b) at low temperatures $T<\mu$.

(ii) The most significant effect of SOC occurs when considering the thermal depletion in case (c), which diverges, i.e., $n_T^{ex}\sim \kappa_x mT \int d\epsilon/\epsilon$. Therefore, {\it the thermal depletion in three dimension becomes infinite in Bogoliubov theory at any finite temperature when both SOC and interaction become isotropic}.  This is a direct consequence of the reduction of $D_{eff}$ to $2$.


It is worthwhile to point out that, while Bogoliubov theory is not expected to be accurate when the depletion becomes close to the total particle number, the above conclusion on the divergent thermal depletion for case (c) is consistent with the general argument that long-range order can not exist in a  three dimensional system with  two soft modes\cite{ PC}. 
When the interaction can be written as $U(n_{\uparrow}+n_{\downarrow})^2$, both the kinetic and interaction energy are constants on the sphere $|{\bf k}|=\kappa_x$. Two soft modes then appear along the tangent directions due to phase fluctuations in the long wavelength limit, whereas the density fluctuation is gapped and suppressed by repulsive interaction in this limit as ordinary condensates. This is similar to Rashba coupling in two dimension where one soft mode exists\cite{Jian}. In our case, the two soft modes in phase fluctuation lead to divergent depletion and the absence of long-range order in case (c).

To verify results (i) and (ii), we perform numerical simulations for the condensate depletion based on Eqs.(\ref{nex_T0}, \ref{nex_T}). We first calculate the dependence of $n^{ex}_{0}/n$ and $n^{ex}_{T}/n$ on the anisotropy of SOC with a fixed value of $\kappa_x\xi(=8)$.  
Fig.(2) shows how $n^{ex}_{0}/n$ and $n^{ex}_{T}/n$ evolve when the configuration of SOC  changes from (a) to (b) by increasing $\kappa_y$ with $\kappa_z=0$ and then from (b) to (c) by increasing $\kappa_z$ with $\kappa_x=\kappa_y$.  One sees that  $n_0^{ex}/n$ increases from $0.15$\% in (a) to $0.5$\% in (b) and finally to $2.1$\% in (c).  For thermal depletion,  $n_T^{ex}/n$ at $T/\mu=0.5$ increases from $0.09$\% in  (a) to $1$\% in (b), and eventually becomes proportional to $-\ln (1-\kappa_z/\kappa_x)$ when $\kappa_z$ approaches $\kappa_x$, as shown by the red dashed line in Fig.(2). We  find that this logarithmical growth of $n_T^{ex}/n$ exists for any finite value of $T$ when SOC approaches the isotropic limit. Therefore, we conclude that $n_T^{ex}/n$ is divergent at any finite temperature in (c). To further confirm the scaling forms in Fig.(1), we also calculate the dependence of condensate depletion in case (a-c) on dimensionless numbers in Eq.(\ref{sca}). As shown in the insets of Fig.(2), numerical results directly verify these scaling forms. 

We note that $n_T^{ex}/n$ enters the logarithmic growth region when $1-\kappa_z/\kappa_x\leq 10^{-4}$ for the parameters we used. While it might be challenging for current experiment to access this regime, the enhancement of thermal depletion is already visible far before entering this regime. As Fig.2 shows, $n_T^{ex}/n$ increases  to $5\%$ when $\kappa_z/\kappa_x=0.98$, which means  $n_T^{ex}/n$ has been enlarged by 50 times compared with case (a). Therefore, the enhanced depletion should be directly observable in experiments.

\begin{figure}
\includegraphics[width=3.2in]{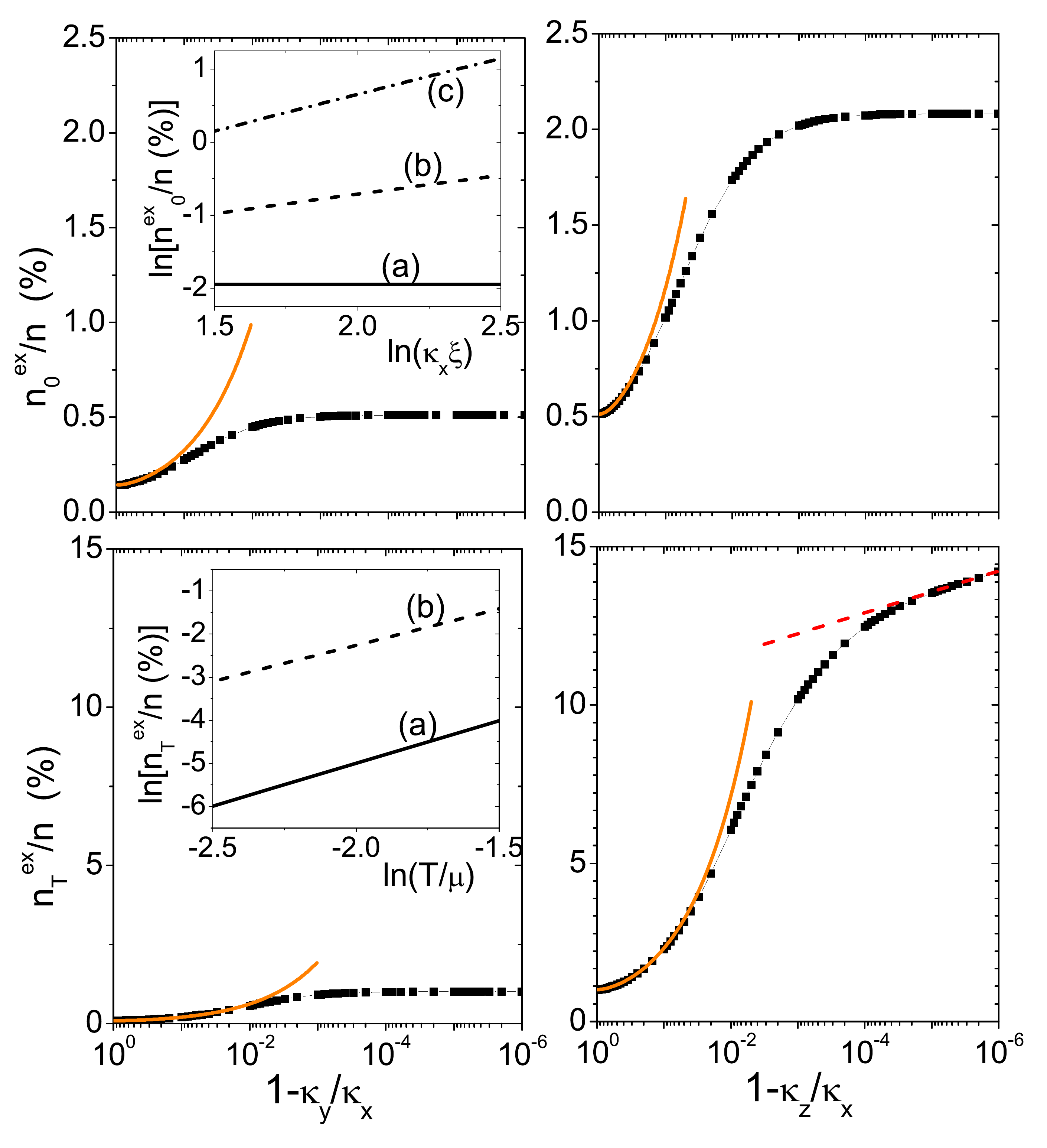}
\caption{(Color Online) $n_0^{ex}/n$ (upper panel) and $n_T^{ex}/n$ at $T/\mu=0.5$ (lower panel)  as functions of 
$1-\kappa_{y}/\kappa_x$ when $\kappa_z=0$(left column) and
$1-\kappa_{z}/\kappa_x$ when $\kappa_x=\kappa_y$(right column), where $U=U_{\uparrow\downarrow}$, $n\tilde{a}^3=10^{-6}$ and $\kappa_x\xi=8$.   Fitting solid curves in orange are based on
$\sim\gamma_{y,z}^{-1/2}$ for
small $\kappa_y$ or $\kappa_z$. Dashed red line is a linear fit $\sim-\ln (1-\kappa_z/\kappa_x)$ when approaching isotropic  limit. 
Inset of the upper panel is a $\ln-\ln$ plot of $n_0^{ex}/n$ as a function of $\kappa_x\xi$ for cases (a, b, c).  Inset of the lower panel is a $\ln-\ln$ plot of $n^{ex}_T/n$ as a function of $T/\mu$ for case (a,b).}
\end{figure}

{\bf Divergent depletion approaching case (c)}.  An important question naturally arises that how the thermal depletion evolves to infinity with increasing $\kappa_z/\kappa_x$ from case (b) to (c), while keeping $U=U_{\uparrow\downarrow}$ fixed? Here we will demonstrate how an energy scale characterizing the anisotropy of SOC determines the depletion.

For $\kappa_z\le \kappa_x=\kappa_y$ and , the expansion of Eq.(\ref{ES}) along the $z$ direction is well approximated by
\begin{equation}
E_{(0,0,q_z)}=\sqrt{\frac{\mu}{m}\left(\gamma_zq_z^2+\frac{q_z^4}{4\kappa_x^2}\right)},\label{Ezs}
\end{equation}
where $\gamma_z=1-{\kappa_z^2}/{\kappa_x^2}$. Here $\gamma_{z}$ reflects the suppression of the sound velocity along the $z$ direction due to a finite value of  $\kappa_{z}$.  We define a characteristic momentum $q_z^*=2\gamma_z^{1/2}\kappa_x$ by setting the contributions from $q_z^2$ and $q_z^4$ terms in Eq.(\ref{Ezs}) to be equal, i.e., $\gamma_zq_z^{*2}=q_z^{*4}/(4\kappa_x^2)$. $q_z^*$ corresponds to an energy scale $E^*_z=2\sqrt{2}\gamma_z c \kappa_x$. $E^*_z\gg \mu>T$ in case (b), while $E^*_z\rightarrow 0$ if case (c) is approached. Whereas both two terms in Eq.(\ref{Ezs}) contribute to quasi-particle DOS, we use the following approximation to obtain the analytic expressions for $n_T^{ex}/n$ in the crossover region from (b) to (c). For $E<E_z^*$, we ignore the contribution from $q_z^4$ term, and replace $\rho(\epsilon)$ by $\rho_<(\epsilon)=\gamma_z^{-1/2}\rho^{(b)}(\epsilon)$. For $E>E_z^*$, we ignore the contribution from $q_z^2$ term, and replace $\rho(\epsilon)$ by $\rho_>(\epsilon)=\rho^{(c)}(\epsilon)$.  Within this approximation, thermal depletion can be written as 
\begin{eqnarray}
n^{ex}_T &\approx& \mu T \int_0^{T} d\epsilon \rho_<(\epsilon)/\epsilon^2,\label{nexb}\\
n^{ex}_T &\approx& \mu T (\int_0^{E^*_z} d\epsilon \rho_<(\epsilon)/\epsilon^2+\int_{E^*_z}^T d\epsilon \rho_>(\epsilon)/\epsilon^2)\label{nexc}
\end{eqnarray}
for $E_z^*>T$ and $E_z^*<T$ respectively. In another word, Eqs.(\ref{nexb},\ref{nexc}) describe $n_T^{ex}$ in the region with large and small value of $\gamma_z$. Note that the above two equations become identical when $E_z^*=T$. To be more explicit, Eqs.(\ref{nexb},\ref{nexc}) give rise to
\begin{eqnarray}
\frac{n^{ex}_T}{n} &=&\beta_1  \gamma_z^{-\frac{1}{2}}(\kappa_x\xi)^{\frac{1}{2}}(n\tilde{a}^3)^{\frac{1}{2}}\left(\frac{T}{\mu}\right)^{\frac{3}{2}} \label{nexsm},\\
\frac{n^{ex}_T}{n}&=&\beta_2 (\kappa_x\xi)({n\tilde{a}^3})\left(\frac{T}{\mu}\right)\left(1+\beta_3\ln \left(\frac{T}{E_z^*}\right)\right), \label{nexla}
\end{eqnarray}
where $\beta_{i=1,2,3}$ are numerical factors.

 Eq.(\ref{nexsm}) tells one that, starting from case (b), thermal depletion increases as a function of  $\gamma_z^{-{1}/{2}}$ for small values of $\kappa_z$.  This behavior of thermal depletion is directly verified by the numerical results in Fig.(2), as demonstrated by the fitting curves in orange.  For a large value of $\kappa_z\lesssim \kappa_x$, Eq.(\ref{nexla}) clearly shows that, at any given temperature, ${n^{ex}_T}$ becomes logarithmically divergent $\sim -\ln E^*_z\sim -\ln(1-\kappa_z/\kappa_x)$ when $\kappa_z$ approaches $\kappa_x$. Therefore, Eqs.(\ref{nexsm}, \ref{nexla}) can be used to qualitatively describe the thermal depletion in the crossover region from (b) to (c).  Similar discussions apply to the condensate depletion in other crossover regions\cite{supple}.

In conclusion, we have pointed out that SOC significantly enhances condensate depletion of bosons, and completely destruct the condensate  when both SOC and interaction approach the isotropic limit. We also show that the dependance of condensate depletion on microscopic variables is distinct for different types of SOC. These results are expected to be useful for future experimental exploration of spin-orbit coupled condensates, and will stimulate more studies on strongly depleted condensates for which correlation effects are important.

We thank T.L Ho, Z. Yu and H. Zhai for helpful discussions. We are particularly grateful to T.L. Ho for valuable comments on the first version of this manuscript which helped us improve the quality of the presentation. XC is supported by the Initiative Scientific Research Program of Tsinghua University and NSFC under Grant No. 11104158. QZ acknowledges startup support from Department of Physics, CUHK.

\section{Supplementary material}

In this supplementary material, we present the results for mean-field ground states, and  analytical expressions of both the quantum and thermal depletion for the crossover regions from (a) to (b), and from (b) to (c).

\subsection{Mean-field ground state}
The kinetic energy shown in Eq.(2) of the main text leads to different manifolds of single-particle ground state, as listed below.
\begin{center}
\begin{tabular}{llll}
(1)&$\kappa_x>\kappa_y,\kappa_z$ &: & two points $\pm{\bf k_0}$, with ${\bf k_0}=\kappa_x{\bf e_x}$\\
(2)& $\kappa_x=\kappa_y>\kappa_z$&: & a circle $k_x^2+k_y^2=\kappa_x^2, k_z=0$ \\
(3)& $\kappa_x=\kappa_y=\kappa_z$&: &a sphere $|{\bf k}|=\kappa_x$\\ 
\end{tabular}
\end{center}
For (1) and (2,3), the single-particle ground states are doubly and infinitely degenerate respectively. The presence of interaction may lift these degeneracy. Consider a plane-wave condensate $|P\rangle_{\bf k}=a_{\bf k}^{\dagger N}|0\rangle$ and a stripe condensate $|S\rangle_{\bf k}=(\frac{a_{\bf k}^{\dagger }+a_{\bf -k}^{\dagger }}{\sqrt{2}})^N|0\rangle$, their interaction energies can be written as
\begin{eqnarray}
&E_{P}({\bf k})=\langle P| H_{int} |P\rangle= \frac{N^2}{2\Omega}\Big(U+2(U_{\uparrow\downarrow}-U)u_{\bf k}^2v_{\bf k}^2\Big)\nonumber\\
&E_{S}({\bf k})=\langle S| H_{int} |S\rangle=\frac{N^2}{4\Omega}\Big((U+U_{\uparrow\downarrow})-2(U_{\uparrow\downarrow}-U)u_{\bf k}^2v_{\bf k}^2\Big).\nonumber
\end{eqnarray}
The expressions for $u_{\bf k}$ and $v_{\bf k}$ have been given in the main text and $u_{\bf k}^2+v_{\bf k}^2=1$. The ground state for (1-3) is determined as follows.

(1) As $u_{\bf \pm k_0}^2=v^2_{\bf \pm k_0}=1/2$, compare $E_{P}({\bf \pm k_0})$ and $E_{S}({\bf k_0})$, we see that the ground state is $|P\rangle_{\bf \pm k_0}$ for $U>U_{\uparrow\downarrow}$, or  $|S\rangle_{\bf k_0}$ for $U<U_{\uparrow\downarrow}$.

(2)The spontaneous symmetry breaking can occur along any direction in the x-y plane. The ground state is $|P\rangle_{\bf k_\phi}$ and $|S\rangle_{\bf k_\phi}$ for $U>U_{\uparrow\downarrow}$ and $U<U_{\uparrow\downarrow}$ respectively, where ${\bf k}_\phi$ is an arbitrary rotation of ${\bf k}_0$ about the $z$ axis.

(3) Despite the high degeneracy of kinetic energy for this case, expressions of $E_{P}({\bf k})$ and $E_{S}({\bf k})$ tell one that interaction energy is minimized when the value of $u^2_{\bf k}v^2_{\bf k}$ is either $1$ or $0$, i.e., when  ${\bf k}={\bf k}_\phi$ or ${\bf k}=\pm{\bf k}_1$ with ${\bf k}_1=\kappa_z{\bf e}_z$. Compare $E_{P}({\bf k_\phi})$, $E_{S}({\bf k_\phi})$, $E_{P}({\bf \pm k_1})$ and $E_{S}({\bf  k_1})$, we find the ground state is $|P\rangle_{\bf k_\phi}$ for $U>U_{\uparrow\downarrow}$, and $|P\rangle_{\bf \pm k_1}$  for $U<U_{\uparrow\downarrow}$. It is worthwhile to mention that though $|S\rangle_{\bf k_1}$ is degenerate with $|P\rangle_{\bf k_\phi}$ for $U>U_{\uparrow\downarrow}$, it is physically unachievable, as spin-up and spin-down particles cannot be mixed with each other at $k_x=k_y=0$.


The results for (1-3) are summarized in the following table. 
\begin{table}[htbp]
\begin{tabular}{|c|c|c|c|}
  \hline
  $ $ & $(1)$ & $(2)$ & $(3)$  \\\hline
  $U>U_{\uparrow\downarrow}$ & $|P\rangle_{\bf \pm k_0}$ &  $|P\rangle_{\bf k_\phi}$ & $|P\rangle_{\bf k_\phi}$ \\\hline
$U<U_{\uparrow\downarrow}$ & $|S\rangle_{\bf k_0}$  &  $|S\rangle_{\bf k_\phi}$ & $|P\rangle_{\bf \pm k_1}$ \\\hline
\end{tabular}
\caption{ Mean-Field Ground state in 3D.}
\end{table}

\subsection{Condensate depletion for the crossover}

\subsubsection{(A) From (a) to (b)}

For $\kappa_z=0$ and an arbitrary $\kappa_y/\kappa_x\in[0,1]$, Eq.(5)(in the main text) at low momentum along the y direction can be expanded as 
\begin{equation}
E_{(0,q_y,0)}=\sqrt{\frac{ \mu}{m}\left(q_y^2 \gamma_y+ \frac{q_y^4}{4\kappa_x^2} f(\frac{\kappa_y}{\kappa_x}) \right)} \label{Ey}
\end{equation}
where 
\begin{equation}
\gamma_y=1-\frac{\kappa_y^2}{\kappa_x^2};
\end{equation}
\begin{equation}
f(x)= x^4+\frac{2E_{\kappa}}{\mu}(1-x^2)^2-\frac{U_{\uparrow\downarrow}}{\mu}x^2(1-x^2).  \end{equation}

For small $\kappa_y/\kappa_x\ll 1$, Eq.\ref{Ey} is dominated by the linear mode, and the sound velocity along the y direction is reduced by a factor of $\gamma_y^{-1/2}$. If $\kappa_y/\kappa_x\rightarrow 1$, Eq.\ref{Ey} is dominated by the quadratic mode. As $\mu\ll E_{\kappa}$, the crossover from the linear mode dominated to quadratic mode dominated regimes occurs when $x=\kappa_y/\kappa_x\rightarrow 1$, and $f(x)\rightarrow 1$. In this limit, by setting $\gamma_yq_y^{*2}=q_y^{*4}/(4\kappa_x^2)$, we define a momentum scale to characterize the crossover, $q_y^*=2\gamma_y^{1/2}\kappa_x$, and the corresponding energy scale is $E^*_y=E_{(0,q_y^*,0)}=2\sqrt{2}\gamma_y c \kappa_x$. 

Similar to the discussions in the main text,  we ignore the $q_y^4$ term in Eq.(\ref{Ey}) for $E<E_y^*$ in calculating the analytic form of quasi-particle DOS, and obtain $\rho_<(\epsilon)= \gamma_y^{-1/2}\rho^{(a)}(\epsilon)$. In contrast, for $E>E_y^*$, we ignore $q^2_y$ and obtain $\rho_>(\epsilon)=\rho^{(b)}(\epsilon)$. Within this approximation, quantum depletion can be written as $n^{ex}_0\approx \mu \int_0^{\mu} d\epsilon \rho_<(\epsilon)/\epsilon$ and $n^{ex}_0\approx \mu (\int_0^{E^*_y} d\epsilon \rho_<(\epsilon)/\epsilon+\int_{E^*_y}^\mu d\epsilon \rho_>(\epsilon)/\epsilon)$  for $\mu<E_y^*$ and $\mu>E_y^*$ respectively.  To be explicit, we obtain
\begin{eqnarray}
&{n^{ex}_0}/{n}=\alpha_1  \gamma_y^{\frac{1}{2}}(n\tilde{a}^3)^{\frac{1}{2}}\label{sky},\\
&{n^{ex}_0}/{n}=\alpha_2 {(\kappa_x\xi)^{\frac{1}{2}}}(n\tilde{a}^3)^{\frac{1}{2}}\left(1-\alpha_3({E_y^*}/{\mu})^{\frac{3}{2}}\right)\label{1DtoR},
\end{eqnarray}
for these two cases, where $\alpha_{i=1,2,3}$ are numerical factors. When $E_y^*$ decreases down to zero at $\kappa_y=\kappa_x$, Eq.(\ref{1DtoR}) becomes the expression for Rashba coupling, i.e, ${n^{ex}_0}/{n}\sim {(\kappa_x\xi)^{\frac{1}{2}}}(n\tilde{a}^3)^{\frac{1}{2}}$, which is consistent with a numerical result found in  Ref.\cite{Ozawa}. 

The same discussions apply to thermal depletion for this crossover, which can be written as $n^{ex}_T\approx \mu T\int _0^T d \epsilon\rho_<(\epsilon)/\epsilon^2$ and $n^{ex}_T\approx \mu T(\int_0^{E_y^*}  d\epsilon \rho_<(\epsilon)/\epsilon^2+\int_{E_y^*}^{T}  d\epsilon \rho_>(\epsilon)/\epsilon^2)$ for $E_y^*>T$ and $E_y^*<T$ respectively, which give rise to
\begin{eqnarray}
&{n^{ex}_T}/{n}= \alpha_1'\gamma_y^{-\frac{1}{2}} (n\tilde{a}^3)^{-\frac{1}{2}}(T/\mu)^2,\\
&{{n^{ex}_T}}/{n}=\alpha_2'  ({ T}/{\mu})^{\frac{3}{2}}{(\kappa_x \xi)^{\frac{1}{2}}}({n\tilde{a}^3}) \left(1-\alpha_3'(E_{y}^*/T ) \right),
\end{eqnarray}
where $\alpha'_{i=1,2,3}$ are numerical factors.

\subsubsection{(B) From (b) to (c)}

For $\kappa_y=\kappa_x$ and an arbitrary $\kappa_z/\kappa_x\in[0,1]$, Eq.(5)(in the main text) at low momentum along z-direction can be expanded as 
\begin{equation}
E_{(0,0,q_z)}=\sqrt{\frac{ \mu}{m}\left(q_z^2 \gamma_z+ \frac{q_z^4}{4\kappa_x^2}g(\frac{\kappa_z}{\kappa_x}) \right)} \label{Ez}
\end{equation}
where 
\begin{equation}
\gamma_z=1-\frac{\kappa_z^2}{\kappa_x^2}(1-\frac{n(U-U_{\uparrow\downarrow})}{4E_{\kappa}});
\end{equation}
\begin{eqnarray}
g(x)&=& x^4(1-\frac{n(U-U_{\uparrow\downarrow})}{E_{\kappa}})+\frac{2E_{\kappa}}{\mu}(1-x^2)^2 \nonumber \\
&&-\frac{U_{\uparrow\downarrow}}{\mu}x^2(1-x^2).  
\end{eqnarray}

Similar to previous discussions for the crossover from (a) to (b), here by setting $\gamma_zq_z^{*2}=q_z^{*4}/(4\kappa_x^2)$, we define a characteristic momentum for the crossover from (b) to (c), $q_z^*=2\gamma_z^{1/2}\kappa_x$, and the corresponding energy scale $E^*_z=2\sqrt{2}\gamma_z c \kappa_x$. 

Following the same procedure in previous discussions, quantum depletion can be written as 
\begin{eqnarray}
&{n^{ex}_0}/{n}=\beta_1' \gamma_z^{-\frac{1}{2}} (\kappa_x\xi)^{\frac{1}{2}}(n\tilde{a}^3)^{\frac{1}{2}}\label{skz}\\
&{n^{ex}_0}/{n}=\beta_2'  (\kappa_x\xi)(n\tilde{a}^3)^{\frac{1}{2}}\left(1-\beta_3'({E_z^*}/{\mu})^{\frac{3}{2}}\right) \label{Rtoi}
\end{eqnarray}
for  $E^*_z>\mu$ and $E^*_z<\mu$ respectively, where $\beta'_{i=1,2,3}$ are numerical factors. With decreasing $E_z^*$, the scaling form of the quantum depletion involves from Eq.(\ref{skz}) to Eq.(\ref{Rtoi}) at $E^*_z=\mu$ and eventually becomes the one for case (c) with isotropic SOC and interaction. 

At finite temperatures, the thermal depletion can be written as Eq.(16) and (17) in the main text  for  $E^*_z>T$ and $E^*_z<T$ respectively. At $E_z^*=T$, the two expressions are identical. With further increasing $E_z^*$ down to zero, thermal depletion eventually becomes divergent, in the form of $-\ln(1-\kappa_z/\kappa_x)$ if $U=U_{\uparrow\downarrow}$ as shown in Fig.2 in the main text.


\begin{thebibliography}{99}
\bibitem{RG} Whereas Rashba coupling renormalizes interaction in the very dilute limit with vanishing density\cite{Ozawa1,Goldbart}, this renomalization becomes less important with increasing density\cite{Goldbart}. In the mean-field region, the expression for interaction is the same as that for ordinary bosons\cite{Goldbart}.
\end{thebibliography}
\end{document}